# Novel Data Models for Inter-operable LCA Frameworks

Kourosh Malek[1,2*], Max Dreger[1], Zirui Tang[3], Qingshi Tu[3*]

[1]Virtual Mind Laboratory, Forschungszentrum Juelich GmbH, Juelich, Germany, [2] Blue A.I., Vancouver, BC, Canada, [3] Department of Forestry, Bio-Product Institute, UBC, Vancouver, BC, Canada

## Abstract

Life cycle assessment (LCA) plays a critical role in assessing the environmental impacts of a product, technology, or service throughout its entire life cycle. Nonetheless, many existing LCA tools and methods lack adequate metadata management, which can hinder their further development and wide adoption. In the example of LCA for clean energy technologies, metadata helps monitor data and the environment that holds the integrity of the energy assets and sustainability of the materials sources across their entire value chains. Ontologizing metadata, i.e. a common vocabulary and language to connect multiple data sources, as well as implementing AI-aware data management, can have long-lasting, positive, and accelerating effects along with collecting and utilizing quality data from different sources and across the entire data lifecycle. The integration of ontologies in life cycle assessments has garnered significant attention in recent years. We synthesized the existing literature on ontologies for LCAs, providing insights into this interdisciplinary field's evolution, current state, and future directions. We also proposed the framework for a suitable data model and the workflow thereof to warrant the alignment with existing ontologies, practical frameworks, and industry standards.

## 1 Introduction

Life cycle assessment (LCA) plays a crucial role in evaluating the environmental impacts associated with the entire life cycle of technologies, products or services. LCA provides a holistic approach to evaluate environmental impacts, offering valuable information for decision-making and for identifying opportunities for resource efficiency and optimization. LCA is often applied to inform environmental policies and regulations, as well as to improve the understanding and management of supply chains.

The evolution of LCA studies has been shaped by advancements in methodology, data availability, and a growing awareness of the need for sustainable practices. Research on the development of LCAs began in the 1970s and gained considerable attention in the 1990s, reflecting the growing awareness of the environmental issues related to products and services. Initially, LCA studies were relatively simple, focusing on a limited number of environmental impact categories, such as fossil fuel depletion and climate change. The data used in these studies were often limited and generic, leading to uncertainties in the results. In the 1990s to the 2000s, as environmental concerns grew and industries sought ways to improve sustainability, LCA methodologies became more sophisticated. The International Organization for Standardization (ISO) developed standards (e.g., ISO 14040 and ISO 14044) to provide guidelines for conducting LCA studies. These standards helped create consistency for the methodology and improve the reliability and comparability of LCA results. In addition, advanced modeling techniques and more comprehensive databases have been developed to capture a more accurate representation of real-world scenarios. This holistic approach, often referred to as life cycle sustainability assessment (LCSA), considers the social and economic dimensions alongside environmental impacts, providing a more comprehensive view of a product's sustainability. With this expanding scope and complexity of LCA methodologies, the incorporation of ontologies has become crucial to enhance the consistency and interoperability of LCA data.

In the realm of data science, an ontology serves as a foundational concept that defines the structure of knowledge within a specific domain. It provides a formal representation of the entities, their attributes, and the relationships between them. By establishing a shared understanding of the meaning of data, an ontology facilitates more effective data integration, sharing, and reuse across different systems and disciplines. In data science, ontologies play a crucial role in enabling machines to interpret and process information in a manner that aligns with human understanding, bridging the gap between raw data and actionable insights.[1]

Ontologies have the potential to address the challenges of integrating and interpreting diverse, heterogeneous data sources for LCAs. They also provide several key enhancements of current LCA frameworks, including the following:

- Data harmonization: LCA involves data from various sources, and ontologies help standardize and harmonize the representation of data. This ensures that different stakeholders can use and understand the data consistently.
- Interoperability: Ontologies enable interoperability between different LCA tools and databases, ensuring that data from various domains can be consistently understood and utilized, thus overcoming the limitations of traditional LCA data formats which often only support syntactic interoperability. [2,3] A common semantic framework allows for the seamless exchange of data and information between various platforms, enhancing collaboration and data integration.
- Semantic clarity: Ontologies provide a shared vocabulary and clear definitions for LCA terms, reducing ambiguity and improving the semantic clarity of LCA studies. This is essential for clear and accurate communication and understanding of LCA results.
- Facilitating automation: Ontologies facilitate the automation of LCA processes by providing a structured and standardized framework for data representation. This improves the efficiency and reliability of LCA studies.
- Adaptability to evolving knowledge: LCA is a dynamic field with evolving information and knowledge. Ontologies can be updated to incorporate new data and adapt to changes in the understanding of environmental impacts and sustainability.

Efforts have been made to develop core ontology patterns and semantically enriched databases for LCA. These initiatives aim to standardize the representation of LCA data, enabling better data sharing, management, and reuse across different LCA studies and applications. Despite recent progress, however, a significant remaining challenge is achieving widespread consensus and adoption of a unified ontology framework within the LCA community. In addition, ongoing tasks include integrating ontologies with existing LCA models and databases, as well as continuously refining ontological frameworks to accommodate evolving LCA methodologies and requirements.

Similar to any fields of research, LCA research data management consists of research datasets in machine or human readable forms (docs, text, spreadsheets, lab notebooks, images, models, algorithms, scripts, methods, know-how, workflows, operating procedures and protocols) that unify (storage technologies) activities across different domains. Several methods are utilized in the recent years to accelerate knowledge flow across human and machine-readable formats such as Universal Resource Identification-URI (e.g. DOI, tags), visualization standards (e.g. JSON), resource description framework –RDF (SPARQL, Neo4J), and force matching data graphs-a proprietary interactive knowledge graph with linked grid structure, rendering images and graphs for orchestration at scale.[4]

The objective of this paper is therefore to propose an optimized data model to facilitate the alignment among existing ontologies, frameworks, and industry standards based on a literature review of existing ontologies for LCA. Ensuring consistency in data representation will contribute to the broader goal of creating a unified LCA ecosystem that supports interoperability.

## 2  Data transparency, the FAIR data principal, and LCA model accessibility

FAIRification of meta data is a widely-accepted guiding principle for data management that ensures findability, accessibility, inter-operability and reproducibility of the data across its entire life cycle (see for example FAIRmat[5], NOMAD[6], NFDI[7], GOFAIR[8] in materials sciences). The FAIR principle[9] does not mine, curate or analyze data. Similarly, it does not serve as a blueprint for software tool or data repository applications and does not suggest ontology or metadata standard by itself.

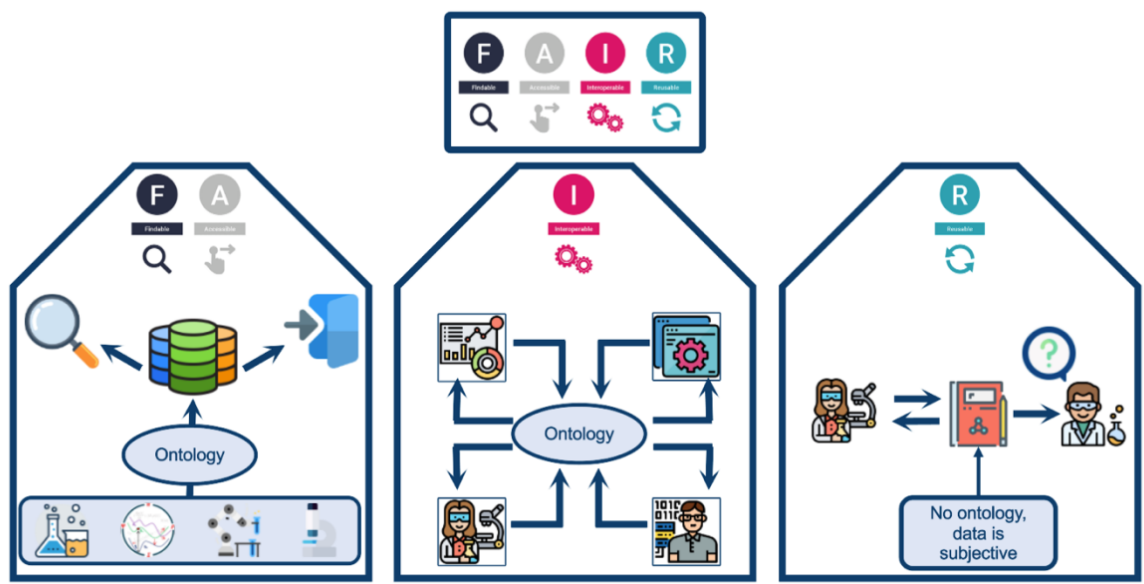

**Figure 1.** Fair metadata management and its relation to ontologies.[1]

AI-ready data schemes require findable, accessible, interoperable, and reusable (FAIR) meta-data standards, i.e., metadata with consistent labels, well-documented data models, and community-accepted ontologies. The processes of generating or labelling training data, as well as the actual training itself need to allow for reproducibility and auditability, for example, to expose data biases. Technical aspects also must be considered thoroughly, including scalability, efficiency of storage, compatibility of data formats with chosen AI methods, as well as data security, ownership, and privacy. High-fidelity and state-of-the-art AI methods and applications therefore need to push data beyond the FAIR principles.

Central for LCA data are ontologies, standardized and community-endorsed languages to describe data.

State-of-the-art data models across domains, however, require collective action to synergize ontologies (Dreger et al., 2023) and identify mappings between ontologies for domain-agnostic data representations. Several AI-assisted approaches for ontologies are currently being developed. By fine-tuning existing generative pre-trained transformer (GPT) models, domain-specific ontologies can be enhanced. Additionally, generalized techniques—rich semantic embeddings along with methods like retrieval augmented generation (RAG)—can be used across research domains to query and extract information mapped to domain-specific ontologies (LLM-enhanced ontology mapping).

## 3  Review of existing ontologies for LCA

Several types of ontologies have been developed to support LCA research. *Environmental impact ontology*[10] focuses on representing concepts related to various environmental impacts, such as climate change, resource depletion, and pollution; it also defines the relationships between different impact categories and their subcategories; *Product life cycle ontology*[11] models the stages and processes involved in the life cycle of a product, from raw material extraction to final disposal; it includes concepts related to manufacturing, transportation, use, and end-of-life considerations; *Material/energy flow ontology*[12] represents the flow of materials/energy through different stages of a product's life cycle; it defines relationships between material/energy inputs, outputs, and transformations.; *Emission factor ontology*[13] defines concepts related to emission factors which quantify the amount of pollutants released per unit of activity; it enables the representation of emissions associated with various processes and activities; *Impact pathway ontology*[14] defines the pathways through which emissions and environmental impacts propagate through ecosystems; it captures the cause-and-effect relationships between emissions and their ultimate environmental consequences; *Social LCA ontology*[15] focuses on representing social aspects and impacts throughout the life cycle, and includes concepts related to human rights, labor conditions, and social well-being; *Data ontology* provides a standardized representation of data used in LCA studies, including data sources, units, and quality attributes; the ontology also aids in harmonizing and integrating data from diverse sources; and *Supply chain ontology*[16]. represents concepts related to supply chain processes, actors, and relationships; it aids in modeling and analyzing the environmental and social impacts associated with the supply chain.

The existing ontologies applicable to LCA and their main characteristics, including the ontology name, components, model types, and applications, are listed in Table 1. LCA-oriented ontologies are primarily utilized for improving data sharing, integration, and interoperability in the field.[17] For example, Kuczenski et al. (2016)[18] argued that discrepancies in formats, interfaces, and flow descriptions across LCA databases (e.g., Ecoinvent, GaBi, and US LCI) might hinder practitioners from effectively sharing and reusing LCA models. Therefore, the authors developed a consensus ontology model to standardize the data mapping process through predefined classes and properties, thereby enhancing the interpretation of data from diverse sources. Mittal et al. (2018)[19] proposed an automated inventory modeling method for chemical manufacturing based on an ontology. The authors emphasized the need for managing data describing both the lineage (or synthesis pathway) and processing conditions of chemical manufacturing, proposing a lineage ontology and a process ontology, respectively, to manage these aspects. They then coupled these two inventory modeling ontologies for chemical manufacturing processes, focusing on the detailed material and energy flows associated with a chemical's supply chain. Meyer et al. (2020)[20] focused on enhancing life cycle chemical exposure assessments through ontology modeling. They develop two ontologies, ExO and LciO, for exposure and life cycle inventory modeling, respectively. The two ontologies were then linked to support integrated life cycle human exposure modeling. Their approach facilitated data sharing and integration between exposure science and LCA, demonstrating the use of ontologies for structured knowledge representation, improved data accessibility, and the promotion of interoperability between diverse data domains within environmental health assessments. Wilde et al. [17] outlined the development of a generic ontology to support various applications in life cycle engineering, addressing the challenges of data interoperability and the integration of life cycle technologies (LCTs). Their work proposes a pathway towards enhancing the analysis and optimization of product life cycles through a structured ontological framework. Saad et al. [21] developed a life cycle inventory knowledge graph (LCIKG) model to represent data of various activities in LCI databases. The authors indicated that the graph database could simplify the querying process for LCI data of a product system compared to the traditional SQL database. Wang et al.[22] combined the process flow knowledge graph (PFKG) with a recommendation system. This tool can facilitate LCA practitioners in querying and retrieving LCA inventory data by simply entering descriptions of products or activities.

In the reviewed papers, applying resource data framework (RDF) was the most common way of building and representing ontologies. RDF uses a triple store framework to organize data as a subject, a predicate, and an object. It is a flexible, scalable, and standardized framework for data representation and interchange on the Semantic Web. Besides, some recent studies have utilized the labeled property graph (LPG) model to manage query-linked data.[21,22] This data framework uses nodes to represent entities and edges to illustrate relations within the specified domain. Compared to RDF, the LPG model offers a more compact structure that can effectively minimize the size of the knowledge graph.[21]

**Table 1:** Summary of LCA-oriented ontologies

| Ontology | Components | Model types | Performance | Application | Citation |
|---|---|---|---|---|---|
| Life cycle engineering (LCE)-based ontology | The ontology employs a range of classes to represent core concepts of LCE, including materials, products, locations, and life cycle technologies (LCTs). Each class encompasses three basic properties: technical, ecological, and economic. Instances comprise objects that occur during the LCE, such as glass fiber, plastic, and steel, as objects in the material class. | Not mentioned | Validation: 1 Scalability: 1 Availability: 1 | The ontology enhances structured data and knowledge for life cycle engineering and can facilitate targeted data acquisition through a standardized framework. | Wilde et al., 2022[17] |
| LCA-oriented process flow ontology | The ontology encompasses the nodes representing the main components of LCA modeling, such as Process, Flow, FlowProperty and Inventory. These nodes are connected by the edges (or relationships). For example, 'hasInputFlow' and 'hasOutputFlow' are defined to link the node Process to the node Flow. | Labeled property graph (LPG) | Validation: 1 Scalability: 1 Availability: 1 | It can support life cycle inventory analysis and facilitate LCA practitioners in searching LCI data through a recommendation system. | Wang et al., 2022[22] |

| Name | Description | Format | Evaluation | Purpose | Reference |
|---|---|---|---|---|---|
| LCA inventory ontology | The study includes three types of ontology corresponding to three Ecoinvent data structures. For example, the ontology for the unit process (UPR) dataset consists of nodes (e.g., UnitProcess, IntermediateFlow, and ReferenceProduct) and edges (e.g., 'Has_flow' and 'Has_values') that represent the main concepts of the dataset. | Labeled property graph (LPG) | Validation: 2 Scalability: 1 Availability: 2 | The ontology provides graph-based data frameworks to manage and query complex life cycle inventory data. | Saad et al., 2023[21] |
| BONSAI ontology for life cycle sustainability assessment (LCAS) | The ontology includes classes like flow, activity, and agent that represent the main elements in LCAs. Predicates define relationships between entities, such as 'performs,' 'isInputOf,' and 'hasObjectType.' | Resource Data Framework (RDF) | Validation: 2 Scalability: 2 Availability: 2 | The ontology enhances the integration, sharing, and reuse of LCAS data across various datasets and platforms (e.g., EXIOBASE, YSTAFDB). | Ghose et al., 2022[2] |
| LCA-oriented ontology for product manufacturing | The product life cycle ontology includes flow and process ontologies and their associated properties. The flow ontology classifies substances into elementary, product, and waste flows, marked by quantitative and substance properties such as 'hasQuantity' and 'hasFormula.' The process ontology outlines the life cycle activities and is segmented into four classes: production, transportation, usage, and disposal phases. It connects to the flow ontology through relation properties like 'hasInflow' and 'hasOutflow.' Besides, the boundary property (e.g., 'hasBoundary') of the process ontology specifies the system boundaries of the product life cycle. | Resource Data Framework (RDF) | Validation: 1 Scalability: 1 Availability: 2 | The ontology provides an interoperable and scalable model that can represent the LCA information for the product life cycle (e.g., processes, flows, and semantic relationships). Besides, it enables semantic queries for life cycle inventory data. | Zhang et al., 2015[3] |
| Consensus ontology-based model for LCA | The ontology categorizes three fundamental components of LCA into classes: activities, flows, and flow quantities. Entities within these classes can engage in two primary types of relationships: exchange and characterization. The exchange relationship connects an activity instance and a flow instance by 'hasInputs' and 'hasOutputs'. By contrast, characterization defines the connection between a flow and a flow quantity, detailing the quantitative attributes of the flow like 'hasMany'. | Resource Data Framework (RDF) | Validation: 2 Scalability: 2 Availability: 1 | The ontology model provides a structured and consensus framework to enhance data interoperability in LCA, facilitating the interpretation of data from varied sources such as Ecoinvent, ELCD, and GaBi pro. | Kuczenski et al., 2016[18] |

| Life cycle inventory ontology (LciO) | This ontology is modified from the existing 'lcaMin' ontology. It uses subclasses and OWL restrictions to define three types of LCA flow: elementary, intermediate, and product flow. Additionally, it introduces a new class named 'technosphere flow' to represent all non-elementary flows. The predicates include conventional 'Input_Of' and 'Output_Of' and generalized predicates 'has_Destination' and 'has_source' to link the flows with activities and products. Besides, a 'Context' class and the predicate 'contains_Activity' is added to facilitate the reuse of LCI inventories. | Resource Data Framework (RDF) | Validation: 1 Scalability: 1 Availability: 1 | The LciO ontology can be connected with the Exposure Ontology (ExO) to enhance data sharing and integration throughout the chemical and product life cycle. It can also support the automated generation and analysis of LCI models, enabling more efficient handling of LCI data. | Meyer et al., 2020[20] |
|---|---|---|---|---|---|
| Note: The validation score indicates the dataset size used for testing the ontology—'1' means testing with small datasets, while '2' indicates testing with large datasets. The scalability score reflects the ontology's applicability—'1' for industry-specific ontologies and '2' for general models. The availability score is based on the provision of resources—'1' if only provides the ontology framework without supplying the code (e.g., OWL), and '2' if provides both the framework and corresponding code. | | | | | |

We also evaluated the current LCA-based ontologies in terms of validation, scalability, and availability. Regarding validation, three out of the seven studies reported using large datasets to validate their ontologies. For instance, Ghose et al. (2022) developed a consensus ontology to integrate LCI databases with the Semantic Web. They employed the YSTAFDB and EXIOBASE datasets for testing, demonstrating that the ontology could accurately represent the flow systems of each dataset. However, other studies only verified their ontologies with limited datasets. For instance, Wang et al. (2022) tested their process flow ontology using a case of aluminum die-casting process. A more thorough evaluation is necessary to confirm its applicability across diverse scenarios. As for scalability, some ontologies provide a general framework for representing LCA domains, whereas others are tailored with industry-specific domains (e.g., manufacturing, chemicals) to apply LCA ontology. Moreover, we found that only three of the seven studies presented their ontology models by supplying the corresponding code (e.g., OWL files). Providing these resources is crucial since it could significantly improve the reproducibility and reusability of the ontology models.

## 4 Identification of gaps

There is a lack of standardization of ontologies for LCA due to diverse domains and applications, continual evolution of LCA methodologies, heterogeneity in data sources, lack of universal terminology, complexity of environmental systems, and technological and computational challenges. Implementing and maintaining ontologies in LCA studies can require sophisticated technological infrastructure and computational tools. The lack of standardized tools and platforms for ontology development and integration can hinder widespread adoption. Efforts to address the lack of standardization in LCA ontologies include the development of guidelines by organizations such as ISO, but achieving widespread adoption and agreement within the LCA community remains a work in progress. Establishing common frameworks, fostering collaboration, and promoting consensus on terminology are essential steps toward enhancing the standardization of ontologies in LCA studies.

*Multiple Interoperability issues* persist with ontologies for LCA. First, the LCA domain involves a wide range of topics, and different researchers or organizations may develop ontologies that are specific to their needs. The existence of multiple, domain-specific ontologies can hinder interoperability, as it may be challenging to reconcile and integrate diverse ontologies into a cohesive framework. Second, as noted previously, there may be a lack of standardized conventions for representing LCA information in ontologies. The absence of agreed-upon standards can lead to inconsistencies and difficulties in mapping and aligning ontologies developed by different entities. Third, different ontologies may use similar terms with different meanings or different terms with similar meanings. This semantic heterogeneity can lead to misunderstandings and misinterpretations when integrating data from various sources, affecting the overall interoperability of LCA ontologies. Finally, updates, revisions, or modifications to existing ontologies are implemented to accommodate new knowledge or changing requirements. This evolution can introduce compatibility issues, especially when integrating data across systems that use different versions of ontologies.

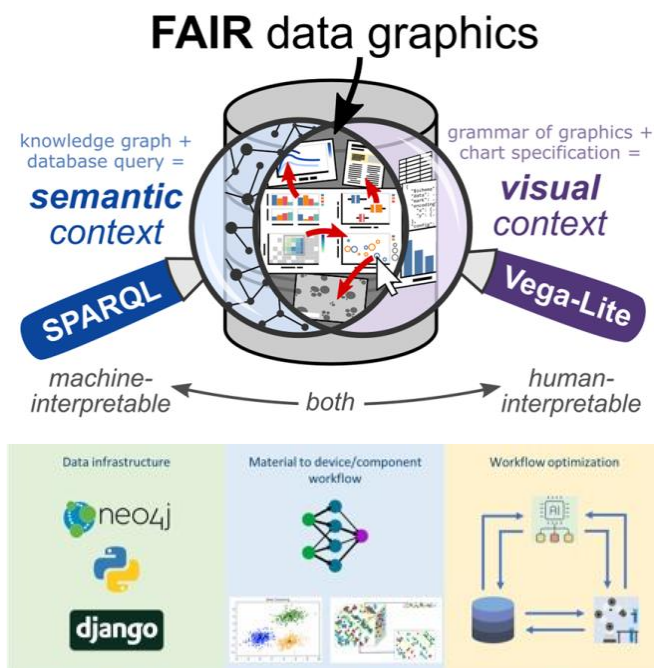

**Figure 1.** (a) Graph data management[23] and (b) the relationship for ontologies and semantic search schemas.[24] Reproduced with permissions. Top image is reproduced with permissions under creative common license (*http://creativecommons.org/licenses/by/4.0/*)[25]

## 5   Overcoming gaps

Traditional data management often entails research teams crafting bespoke data infrastructure for specific projects, leading to isolated datasets across different groups. However, for effective machine learning (ML) applications, larger, more integrated datasets are crucial. The creation of such unified datasets requires merging these disparate datasets, a task that poses significant challenges owing to the lack of standardization. This is where FAIR principles and ontologies have become indispensable by providing guidelines for data management that can bridge the divide between isolated datasets[5-8]. They enhance scientific reproducibility and accessibility to curated data and facilitate the use of data-driven methods while promoting standardization, which is crucial for data integration.

As formal knowledge representations, ontologies are essential for making varied metadata machine readable. Using classes, properties, relationships, and rules, they standardize knowledge representation within specific fields, facilitating data and knowledge exchange among researchers. In the interdisciplinary realm of materials science, standardization through ontologies is crucial. A notable example is the European Materials Modeling Ontology (EMMO)[25], which is a comprehensive ontology developed by the European Materials Modeling Council. EMMO's structure comprises three levels to define real-world objects, specific perspectives, and domain-specific ontologies. It uniquely represents the fabrication, characterization, and simulation of materials and devices. Other projects, such as NOMAD, CHAMEO, and BigMap[10], use EMMO-based ontologies to enhance their creation process and improve interoperability, owing to shared structures and rule sets. Similar to languages, the power of an ontology is tied to its adoption level, and EMMO provides a widely accepted foundation for many materials science domains.

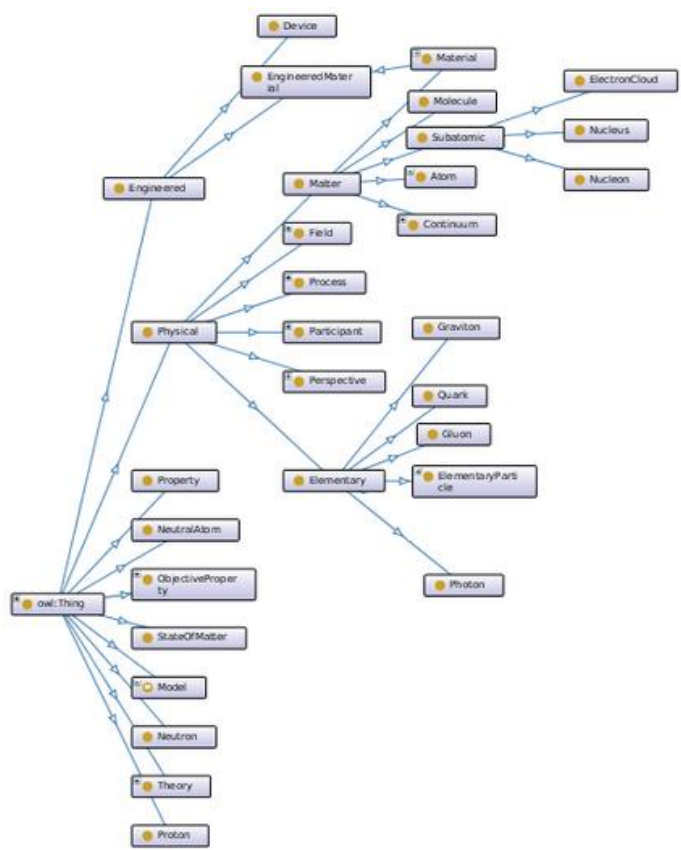

**Figure 2.** EMMO ontology for materials fabrication, modeling and characterizations.[26] Reproduced with permission.

At the heart of data infrastructure is the storing of fabrication, simulation, and measurement data alongside comprehensive metadata for reproducibility. Currently, the industry standard is relational databases such as MySQL or Postgres, which use tables to represent real-world objects and their instances. However, their structured approach can be rigid, and processing relationships using foreign keys can be slow. NoSQL databases offer a more dynamic data model, including document-oriented, key-value, and graph databases. Document-oriented databases store data in various encodings, allowing for a more flexible model, but they still are inefficient at handling relationships. Key-value databases offer flexibility but may require extensions for complex use cases. In this study, we used Neo4j, which is a native graph database that uses graph theory for data representation and storage. This database handles complex domains efficiently, eliminating layers of abstraction between the real world and the data model. The nodes and relationships in Neo4j can have properties, and directed relationships offer more context. Graph databases are particularly beneficial in materials research due to the heterogeneity of data. For example, fabrication workflows can be represented naturally with the use of graphs. Unlike table-based processes, graph databases store workflows as node sequences, making variations easy to accommodate without altering existing data. This flexibility allows for nonprescriptive process structuring, which is crucial in the constantly evolving field of materials science.

Our proposed data model utilizes node sequences with labels corresponding to Python classes defined in the neomodel and within the ontology, providing a structure within the database. Nodes can carry multiple labels that signify specific entities with attributes held as key-value pairs to identify them. Label indexing can accelerate node retrieval, but it may reduce write efficiency and increase storage requirements. In materials science, the attribute space for materials, components, devices, and processes is vast and increasing. To manage this, physical quantities are stored in external properties and parameter nodes, facilitating the efficient querying of similar properties or fabrication parameters. This externalization allows the storage of data as scalar or scalar arrays. Neo4j's data structure enables the efficient traversal of relationships with O(1) complexity. Adjacent nodes are physically stored close together to increase the query efficiency, with the database architecture optimized for frequent queries. Additionally, caching during queries enhances the efficiency of the reading, writing, and matching operations.

Model dockers must be built to certain standards with proper documentation and associated architectures. Metadata standards can be implemented using a graph database management module that matches datasets to specific ontologies[2]. The seamless integration of model management with other orchestration units ensures that the correct information is sent to the data control unit to schedule or execute proper models for data analytics, inverse design, or further device integration.

***Harmonization of ontologies*** for LCA involves aligning and reconciling diverse ontologies within the LCA domain to promote interoperability, consistency, and a shared understanding of concepts. The goal is to create a unified and standardized framework that allows for seamless integration and exchange of LCA data across different systems, tools, and applications.
The first step in harmonization is to identify and catalog existing ontologies relevant to LCA. This includes ontologies developed by different organizations, researchers, or industry initiatives. A comprehensive understanding of the landscape is crucial for effective harmonization. Next, a thorough analysis is required of the structures, concepts, and relationships within identified ontologies to identify commonalities and differences, paying attention to variations in terminology, definitions, and semantic relationships. An integrated ontology framework can then be generated that synthesizes and harmonizes the identified ontologies. This involves creating a unified structure that captures the essential elements of LCA while addressing semantic heterogeneity and ensuring consistency across different ontologies. Terminology should also be harmonized by establishing a common vocabulary across ontologies, defining clear and unambiguous terms, ensuring that similar concepts are represented consistently, and addressing any variations or discrepancies in the naming conventions used. Finally, the harmonized ontology should be validated through practical use cases and applications, applying the integrated framework to real-world

LCA scenarios to assess its effectiveness in facilitating interoperability, data exchange, and consistency across different platforms.

***Graph databases*** can be used to represent and query ontologies for LCA flexibly and efficiently. Graph databases model data as interconnected nodes and edges, making them particularly suitable for capturing complex relationships within LCA ontologies.

In a graph database ontology for LCA, different entities (e.g., processes, materials, impact categories) are represented as nodes, and the relationships between these entities are represented as edges. This graph structure allows for a more intuitive representation of complex relationships in the LCA domain. Nodes in the graph can represent various processes involved in the life cycle of a product or system and can have attributes such as process name, location, and inputs/outputs. Edges between process nodes can indicate relationships, such as the flow of materials or energy between processes. Materials and resources can be represented as nodes in the graph, and edges can indicate the flow of materials between different processes. Transformations, such as the conversion of raw materials into products, can be represented by directed edges between nodes. Impact categories, representing environmental or social impacts, can be modeled as nodes in the graph. Edges can connect these nodes to relevant processes or materials, indicating the contribution of each process or material to specific impact categories. Nodes can represent functional units, defining the reference unit for the LCA study. Edges connect functional units to processes, materials, and impact categories, linking the study's scope to specific elements within the LCA ontology. Graph databases allow for the representation of temporal and spatial relationships. Nodes and edges can be annotated with information about when and where processes occur, providing a more comprehensive view of the life cycle dynamics.

Graph databases can incorporate uncertainty and sensitivity analysis by attaching probabilistic information to nodes or edges. This allows for the representation of variations in LCA data, supporting more robust decision-making processes. These databases can also facilitate ontology evolution and versioning by allowing the addition, modification, or removal of nodes and edges over time. This capability is crucial for accommodating updates in LCA methodologies or incorporating new knowledge into the ontology. They can also provide efficient query capabilities and traversal algorithms, enabling users to explore and analyze complex relationships within the LCA ontology. This makes it easier to retrieve specific information, navigate the ontology, and perform advanced analyses. Finally, graph databases can be integrated with existing LCA tools, databases, or applications, providing a seamless way to connect and exchange data. This integration enhances the overall interoperability of LCA systems.

The ***data quality*** of ontologies for LCA is crucial for ensuring the reliability and effectiveness of environmental impact assessments. The quality of data in LCA ontologies impacts the accuracy, consistency, and trustworthiness of the results obtained. The data quality of LCA ontologies can be assessed based on several parameters:

- Accuracy: The definitions of terms within the ontology should be clear, precise, and accurate, minimizing ambiguity. Ontological representations should accurately reflect the real-world entities, processes, and relationships within the LCA domain.
- Completeness: The ontology should cover a comprehensive range of LCA aspects, including raw material extraction, production, distribution, use, and end-of-life stages. Detailed

- information about processes, materials, and impact categories should be included to ensure a thorough representation of the life cycle.
- Consistency: The ontology should maintain internal consistency, ensuring that terms and relationships are coherent and do not contradict each other. The ontology should align with established standards and conventions within the LCA community to ensure compatibility with existing methodologies and frameworks.
- Precision: Terms should be defined at an appropriate level of granularity to capture detailed information without unnecessary complexity. Where possible, quantitative data (e.g., emission factors, material flow data) should be incorporated with appropriate precision to enhance the accuracy of LCA assessments.
- Traceability: Sources and references for data and assumptions used within the ontology should be documented.

**Role of GenAI: LLM-enhanced and multi-domain data management**

Recent application and high-velocity advancements in LLMs has shown a lasting and strong impact in deployment of intelligent data pipelines for AI/ML models at scale. One such example is the development of LLM-enabled methods for autonomous data extraction from various sources and across various scientific domain to generate multi-modal data assets. The main challenges in this context include the veracity and variety of the extracted data assessment which are required for training novel AI/ML models, or utilized or fine-tuning, testing and validation of pre-trained models. The LLM-based semantic search and autonomous ontology creation thereof, as well as mapping of diverse sets of data assets can greatly overcome these challenges. In particular, the utility of such approaches results in adoption and deployment of a harmonized data model for any MLOps process (machine learning operations) across many scientific domains and disciplines. In order to enhance the domain-specific ontologies, which are pivotal in structuring and interpreting complex LCA datasets, fine-tuning existing generative pre-trained transformer (GPT) models are required. Generalized techniques, such as rich semantic embeddings along with methods like retrieval augmented generation (RAG), can be used seamlessly across various research domains or systems of (data+model+experts) to intelligently offer queries and to extract information, mapped to domain-specific ontologies.

## 6 Use cases

We developed a method to aggregate LCA inventory data on bio-based chemical production processes from diverse publications. These data are mainly encapsulated in tables structured according to LCA ontology principles, showing only minor variations. To standardize and analyze these data, we created a set of unified table structures that can represent the full bandwidth of the data in scientific literature. We manually consolidated two sample tables from the literature into the unified table structure while integrating additional details, such as author information and workflow specifics, from the corresponding publications. The full pipeline is depicted in Figure 3. The data assets are gathered from scientific literature and mapped on the unified tables.

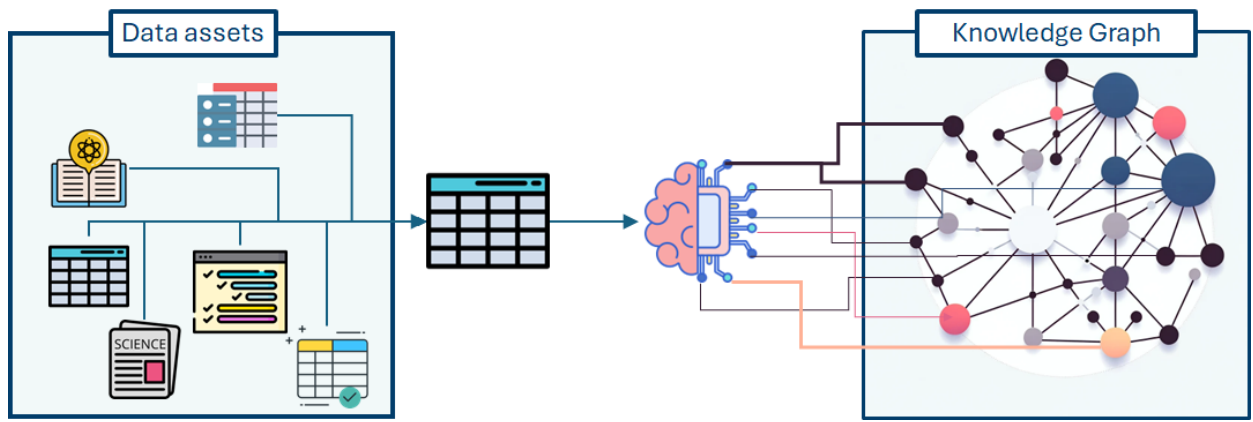

**Figure 3.** Schema of the data ingestion pipeline. Semi-structured data from literature enriched and extracted into a set of unified tables. These tables are input to a cypher function that populates a neo4j knowledge graph.

This unified tables contains the following sub-tables:

- Workflow table (sequences of activity)
- Metadata table (geographic context, etc.)
- Agent table (deployed agents and their features)
- Reference table (author, title etc.)

The workflows described within these tables reference each other via an internal identifier. Structuring the unifier tables in this way keeps the data transformation modular, as adding new tables that can represent different aspects of the workflows becomes straightforward. For each unified table a cypher query was generated that ingests the table into the neo4j database and adds it to corresponding existing data. The actual data ingestion also benefits from the high degree of modularity as newly added tables strictly require the addition of a new cypher function call.

This manual consolidation process will be transformed into an NLP-enabled (semi-) automated extraction pipeline in the future, to automate data extraction and analysis at large scales. The creation of a single import script for ingesting data into Neo4j laid the groundwork for an initial LCA data-management tool. The modularity facilitates automating the data extraction process immensely, as only the transformation of the data assets into unified tables requires dynamic adoption to varying input data.

The unified table structure and associated cypher queries are accessible in the supplementary information. A screenshot of the partially ingested data is shown in Figure *4*, demonstrating the application of the tool.

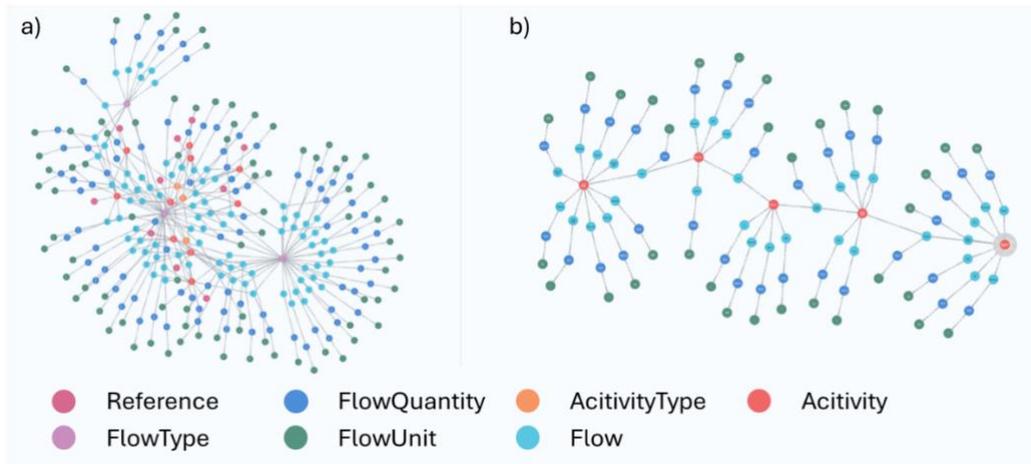

**Figure 4.** Screenshots of the created knowledge graph. A) Workflows extracted from two publications; b) a specific workflow.

The developed database significantly advances domain-specific inquiry capabilities, enabling users to filter through various flows, agents, and regions, and efficiently compare the costs and yields of different workflows. Querying within the database is presently achieved through cypher queries, as examples shown in Figure 5. Future enhancements will focus on integrating NLP technologies to develop a user interface that simplifies the data analysis and retrieval. This interface allows users to interact with a database without direct access to the database, thereby reducing complexity and increasing security for data analysis tasks. This strategic direction not only streamlines the user experience, but also broadens the accessibility of LCA data for comprehensive analysis and decision-making in bio-based chemical production.

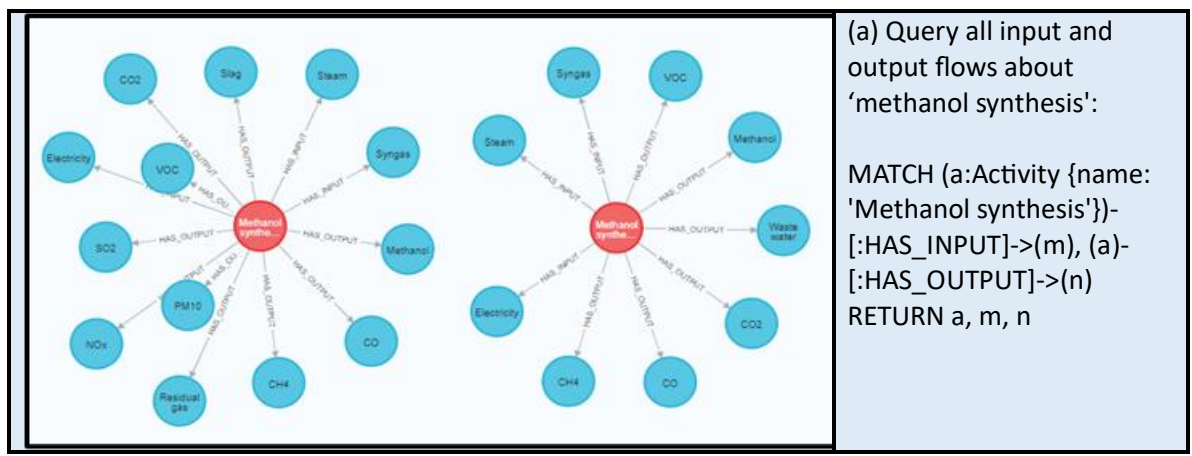

(a) Query all input and output flows about 'methanol synthesis':

MATCH (a:Activity {name: 'Methanol synthesis'})-[:HAS_INPUT]->(m), (a)-[:HAS_OUTPUT]->(n) RETURN a, m, n

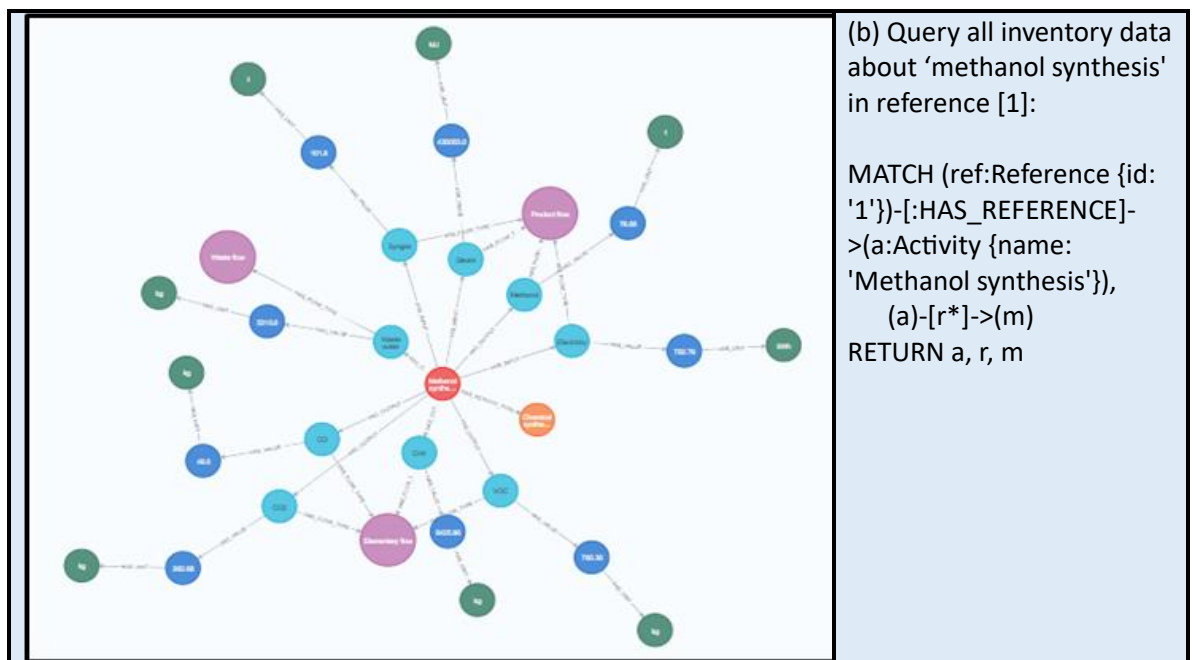

**Figure 5.** Query examples for the LCA dataset.

## 7   Conclusions

The integration of ontologies in LCA methodologies has garnered significant attention in recent years, reflecting the growing need for efficient data management and analysis in the field of sustainability. This comprehensive review synthesized the existing literature on ontologies for LCAs, providing insights into this interdisciplinary field's evolution, current state, and future directions. We found a lack of standardization of ontologies for LCA due to a variety of reasons, including diverse domains and applications, continual evolution of LCA methodologies, heterogeneity in data sources, lack of universal terminology, complexity of environmental systems, and technological and computational challenges. Implementing and maintaining ontologies in LCA studies can require sophisticated technological infrastructure and computational tools. The lack of standardized tools and platforms for ontology development and integration can hinder widespread adoption. Future efforts to address the lack of standardization in LCA ontologies include the development of guidelines by organizations such as ISO, establishing common frameworks, fostering collaboration, and promoting consensus on terminology. We also recommend further research on the integration of knowledge graph and LLMs to enhance the scalability of data ingestion and query for the ontology-based LCA frameworks.

**Acknowledgment**

The authors acknowledge the support of the Natural Sciences and Engineering Research Council of Canada (NSERC) [funding reference number RGPIN-2021-02841].